\documentclass[]{interact}

\usepackage{diagbox}
\usepackage{makecell}
\usepackage{caption}
\usepackage{graphicx}
\usepackage{float}
\usepackage{subfigure}

\usepackage[natbibapa,nodoi]{apacite}
\usepackage{epstopdf}

\usepackage[longnamesfirst,sort]{natbib}
\bibpunct[, ]{(}{)}{;}{a}{,}{,}
\renewcommand\bibfont{\fontsize{10}{12}\selectfont}

\theoremstyle{plain}

\theoremstyle{definition}

\theoremstyle{remark}

\begin{document}


\title{EmotionBox: a music-element-driven emotional music generation system using Recurrent Neural Network}

\author{
	\name{Kaitong Zheng\textsuperscript{a,b}, Ruijie Meng\textsuperscript{a,b}, Chengshi Zheng\textsuperscript{a,b},
		Xiaodong Li\textsuperscript{a,b}, Jinqiu Sang\textsuperscript{a,b}\thanks{CONTACT Jinqiu Sang Email: sangjinqiu@mail.ioa.ac.cn}, Juanjuan Cai\textsuperscript{c} and Jie 
		Wang\textsuperscript{d,e}}
	\affil{\textsuperscript{a}Key Laboratory of Noise and Vibration Research, Institute of Acoustics, Chinese Academy of Sciences, Beijing, China; 
		\textsuperscript{b}University of Chinese Academy of Sciences, Beijing, China;
		\textsuperscript{c}State Key Laboratory of Media Convergence and Communication, Communication University of China, Beijing, China;
	    \textsuperscript{d}School of Electronics and Communication Engineering, Guangzhou University, Guangzhou, China;
        \textsuperscript{e}National Environmental Protection Engineering and Technology Center for Road Traffic Noise Control, Beijing, China}
}


\maketitle

\begin{abstract}
With the development of deep neural networks, automatic music composition has made great progress. Although emotional music can evoke listeners' different emotions and it is important for artistic expression, only few researches have focused on generating emotional music. This paper presents EmotionBox -an music-element-driven emotional music generator that is capable of composing music given a specific emotion, where this model does not require a music dataset labeled with emotions. Instead, pitch histogram and note density are extracted as features that represent mode and tempo respectively to control music emotions. The subjective listening tests show that the Emotionbox has a more competitive and balanced performance in arousing a specified emotion than the emotion-label-based method.
\end{abstract}


\section{Introduction}\label{section1}

Computational modeling of polyphonic music has been deeply studied for decades \shortcites{Westergaard1959}\citep{Westergaard1959}. Recently, with the development of deep learning, neural network systems for automatic music generation have made great progress on the quality and coherence of music \shortcites{Herremans2019,Jin2020,Herremans2017}\citep{Herremans2019,Jin2020,Herremans2017}. 

As we know, emotion is of great importance in music since music is an excellent mood inducer \shortcites{Raynor1958}\citep{Raynor1958}. Therefore, emotion is an important property when composers write their works. However, surprisingly, automatic systems rarely consider emotion when generating music, which lacks the ability to generate music based on a specific emotion.

To solve this problem, it is necessary to explore the relation between music emotions and music elements.
As mentioned in \shortcites{Parncutt2014}\citep{Parncutt2014}, the relation in Western tonal music between emotional valence (positive versus negative) and music-structural factors such as tempo (fast versus slow) and mode (major versus minor tonality) are well established in music psychology.
Experimental data has illustrated that fast tempo tends to make music sound happy while slow tempo has the opposite effect \shortcites{Rigg1940}\citep{Rigg1940}. 
In typical tonal musical excerpts, the experimental result \shortcites{Gagnon2003}\citep{Gagnon2003} shows that tempo is more determinant than the mode in forming happy-sad judgments.
As mentioned in \shortcites{Cheng2020}\citep{Cheng2020}, the mode is determined by the subset of pitches selected in a certain piece. According to music theory, the interval between the first and third tone is a major third in a major scale whereas it is a minor third in a minor scale. Many experiments have demonstrated that musical excerpts written in the major or minor mode are judged to be positive or negative, respectively \shortcites{Hevner1935,Hevner1936}\citep{Hevner1935,Hevner1936}.

Most previous emotional music generation models are based on emotion labels  \shortcites{Ferreira2019,Zhao2019,ferreira2020computer}\citep{Ferreira2019,Zhao2019,ferreira2020computer}, without taking into consideration the effect of music elements. Moreover, label-based methods require a huge music dataset labeled with different emotions, which often needs a lot of tedious work.
Whether or not we can utilize the relation between music emotions and music elements instead of a large number of emotion labels to supervise the training of the deep neural network and what music elements are suitable for presenting the emotion of music are the main focuses in this paper.

To this end, we extract two features from two music elements (i.e. tempo and mode) to supervise the deep neural network for generating music with a specific emotion. 
To the best of our knowledge, this is the first music-element-driven emotional symbolic music generation system based on a deep neural network.

\section{Related Work}\label{section2}

Currently, deep learning algorithms have become mainstream methods in the field of music generation research. Music generation can be classified into two types: symbol domain generation (i.e. generating MIDIs or piano sheets \shortcites{Yang2017,Dong2018}\citep{Yang2017,Dong2018}) and audio domain generation (i.e. directly generating sound wave \shortcites{Oord2016,Schimbinschi2019,Subramani2020}\citep{Oord2016,Schimbinschi2019,Subramani2020} ). 

Recurrent Neural Network (RNN) or its variants have been widely used to model sequential data. Its outstanding temporal modeling ability makes it suitable for music generation. The first attempt is that Todd used RNN to generate monophonic melodies early in 1989 \shortcites{Todd1989}\citep{Todd1989}. 
To solve the gradient vanishing problem of RNN, Eck et al. proposed an LSTM based model in music generation for the first time\shortcites{Eck2002}\citep{Eck2002}. 
In \shortcites{boulanger2012modeling}\citep{boulanger2012modeling}, RNN combined with Restricted Boltzmann Machines was proposed to model polyphonic music, which is superior to the traditional model in various datasets. 
In 2016, the magenta team proposed the Melody RNN model which can generate long-term structures in songs \shortcites{Todd1989}\shortcites{waite2016generating}\citep{waite2016generating}.
In 2017, Anticipate RNN \shortcites{hadjeres2017interactive}\citep{hadjeres2017interactive} was used to generate music interactively with positional constraints.
Moreover, Bi-axial LSTM (BALSTM) \shortcites{Johnson2017}\citep{Johnson2017} proposed by Johnson et al. is capable of generating polyphonic music while preserving translation invariance of the dataset. 
Recently, more advanced deep generative models such as VAE \shortcites{hadjeres2017interactive,Brunner2018}\citep{hadjeres2017interactive,Brunner2018}, GAN \shortcites{Guan2019,Huang2019}\citep{Guan2019,Huang2019}, and Transformer \shortcites{Huang2019,Zhang2020}\citep{Huang2019,Zhang2020} have gradually been used in music generation.

The expressive generation has been long explored in the field of computer music, reviewed in \shortcites{Kirke2009}\citep{Kirke2009}.
With the development of deep learning, there are several previous attempts to generate emotional music based on deep neural networks.
Ferreira et al. proposed an mLSTM based model that can be directed to compose music with a specific emotion as well as analyze music emotions \shortcites{Ferreira2019}\citep{Ferreira2019}. However, only video game soundtracks are used in training and evaluation. 
In 2019, Zhao et al. extended the BALSTM network proposed in \shortcites{Mao2018}\citep{Mao2018} and used the model in emotional music generation \shortcites{Zhao2019}\citep{Zhao2019}. 
Recently, Ferreira et al. proposed a system called Bardo Composer, which generates music with different emotions for tabletop role-playing games based on the mood of players \shortcites{ferreira2020computer}\citep{ferreira2020computer}.
However, all methods mentioned above are label-based thus a large dataset labeled with emotions is needed. Moreover, to the best of our knowledge, none of the labeled datasets is available online. Labeling the dataset manually takes a lot of time and effort. In our work, we train the model on an open-source MIDI dataset without emotion labels.

\section{Data Preprocessing}

\subsection{Note Representation}

The input of our proposed generation model consists of polyphonic MIDI files, which are composed of both melody and accompaniment. To present notes with expressive timing and dynamics, we use the performance encoding proposed in \shortcites{Oore2020}\citep{Oore2020}, which consists of a vocabulary of NOTE-ON, NOTE-OFF, TIME-SHIFT, and VELOCITY events. The main purpose of encoding is to transform the music information in MIDI files into a suitable presentation for training the neural network.

The pitch information in MIDI files ranges from 0 to 127, which is beyond the pitch range of a piano. In our work, pieces in the training set are all performed by piano. Thus, the pitch range is only presented from 21 to 108, which correspond to A0 and C8 on piano, respectively.
For each note, music dynamics is recorded in MIDI files, ranging from 0 to 127 to present how loud a note is. For convenience, we use velocity ranges from 0 to 32 to convey the dynamics. The range can be mapped to 0 to 127 when generating MIDI files. 

Finally, a MIDI excerpt is represented as a sequence of events from the following vocabulary
of 240 different events:

\begin{itemize}
	\item 88 NOTE-ON events: one for each of the 88 (21-108) MIDI pitches. Each event
	starts a new note.
	\item 88 NOTE-OFF events: one for each of the 88 (21-108) MIDI pitches. Each
	event releases a note.
	\item 32 TIME-SHIFT events: each event moves the time step forward by
	increments of 15ms up to 1 second.
	\item 32 VELOCITY events: each event changes the velocity applied to all upcoming notes.
\end{itemize}

\subsection{Feature Extraction}

In this work, we feed the model with two manually extracted musical features, namely pitch histogram, and note density.
A pitch histogram \shortcites{Tzanetakis2003}\citep{Tzanetakis2003} is an array of 12 integer values indexed by 12 semitones in a chromatic scale, showing the frequency of occurrence of each semitone in a music piece. 
An example of a pitch histogram in C major is shown in Table \ref{table:pitch histogram}.
According to music theory, notes with a sharp sign are not included in C major. 
Therefore, in this work, we set their corresponding value in pitch histogram as 0 so that they will never be played in a C major music.
C, F and G are the tonic, subdominant, and dominant in C major respectively. They are the main elements in a C major music so their corresponding value in pitch histogram is set as 2, which means the probability of starting these notes is twice as much as other notes in C major.
Pitch histograms can capture musical information regarding harmonic features of different scales.

Note density is a number to record how many notes will be played within a time window (2 seconds in our work). Note density can present the speed information in each part of a music piece. Note density and pitch histogram are calculated at each time step.

The motivation for this is that we can explicitly choose a pitch histogram and note density when creating samples, which provides us two options to control the music generation. By changing the pitch histogram and note density, we can therefore alter the mode and tempo of the music, which ultimately leads to emotional difference.

\begin{table}[h!]
	\centering
	\renewcommand\arraystretch{1.5}
	\setlength{\tabcolsep}{4pt}
	\small
	\caption{An example of pitch histogram in C major scale}
	\label{table:pitch histogram}
	\begin{tabular}{ |c | c c c c c c c c c c c c| } 
		\hline
		Pitch name & C & C$^\sharp$ & D & D$^\sharp$ & E & F & F$^\sharp$ & G & G$^\sharp$ &  A & A$^\sharp$ & B \\ 
		\hline
		Pitch histogram & 2 & 0 & 1 & 0 & 1 & 2 & 0 & 2 & 0 & 1 & 0 & 1 \\ 
		\hline
		Probability distribution &  0.2 & 0 & 0.1 & 0 & 0.1 & 0.2 & 0 & 0.2 & 0 & 0.1 & 0 & 0.1 \\ 
		\hline
	\end{tabular}	
\end{table}

\subsection{Russell Emotion Model}

\begin{figure}[h!]
	\centering
	\includegraphics[width = 0.6\textwidth]{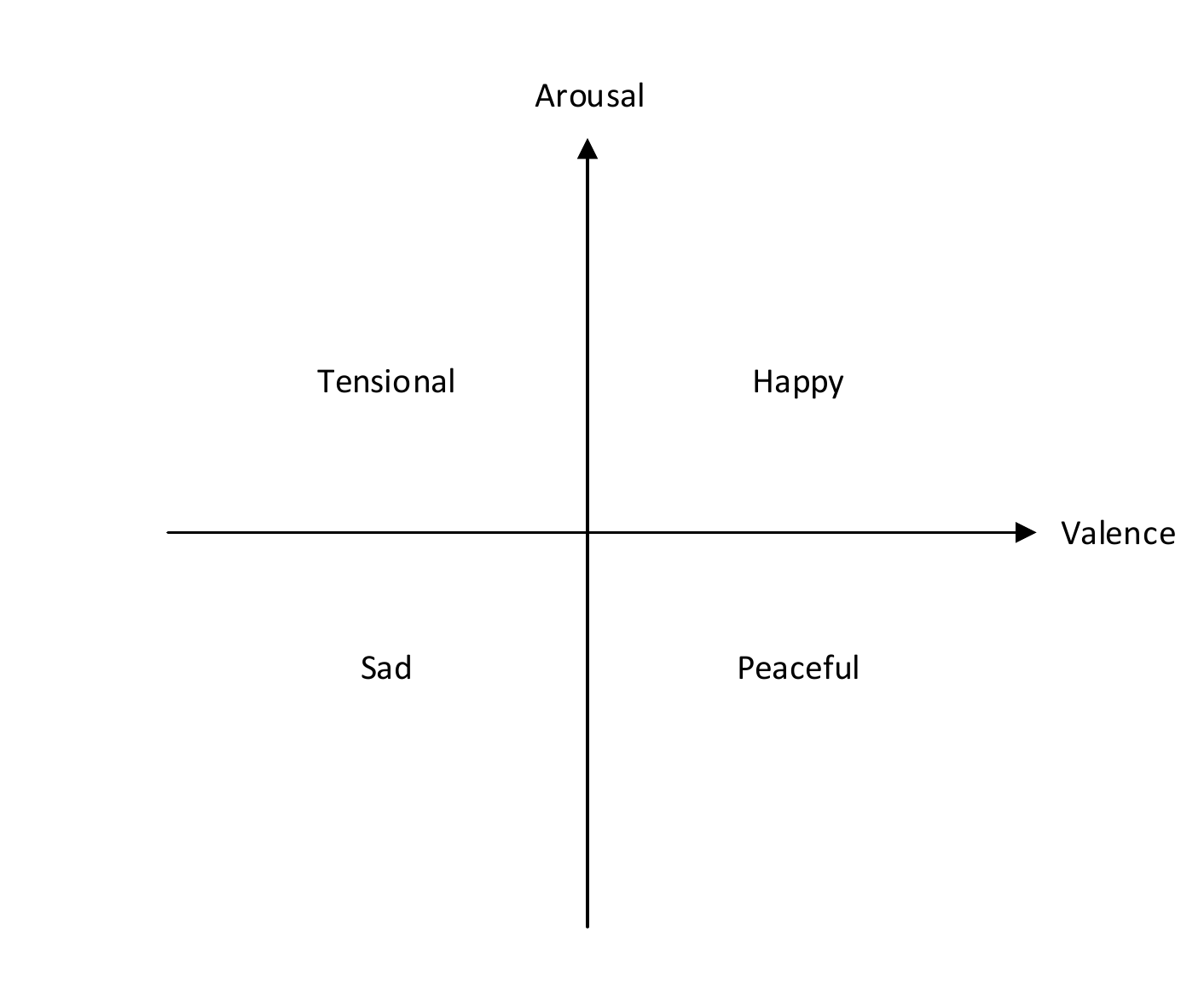}
	\caption{Simplified Russell's two-dimensional Valence-Arousal emotion space. The x-axis denotes valence while the y-axis denotes arousal.}
	\label{pic:Russell}
\end{figure}

Russell proposed a two-dimensional model to describe the affective experience. Two coordinate axes present the degree of pleasure and arousal respectively as shown in Figure \ref{pic:Russell}. For simplicity, we only use four basic emotions as shown in four quadrants. Our model is designed to generate music with these four basic emotions, namely happy, tensional, sad, and peaceful. The four emotions are located in four different quadrants, presenting four varying degrees of valence and arousal. 

\subsection{Emotion Presentation}

As we have mentioned in the introduction, there is a strong connection between music elements and music emotional valence. 
Therefore, we combine note density and pitch histogram to control the tempo and mode of the generated sample. According to twelve-tone equal temperament, an octave is divided into 12 parts, all of which are equal on a logarithmic scale. So, we can choose the mode when generating music by changing the probability of each semitone.
We use an array containing 12 integers to present a pitch histogram. For example, C major is presented as [2, 0, 1, 0, 1, 2, 0, 2, 0, 1, 0, 1] where 2 presents the tonic, subdominant, and dominant while 1 presents other notes in the scale. Pitch histogram of C minor is presented as [2, 0, 1, 1, 0, 2, 0, 2, 1, 0, 1, 0] according to music theory. A pitch histogram is used to control the valence of music.

Note density indicates the number of notes that will be performed within 2 seconds (the time window is  adjustable). We set note density as 1 to present slow music and note density as 5 to present fast music. Note density is used to control the arousal of music.
Combining mode and note density as two adjustable parameters, we aim to generate four categories of emotional music: happy (with the major scale and fast tempo), tensional (with the minor scale and fast tempo), peaceful (with the major scale and slow tempo), sad (with the minor scale and slow tempo).

\section{Method}\label{section3}

\subsection{Neural Network Architecture}

\begin{figure}[h!]
	\centering
	\includegraphics[width = 0.95\textwidth]{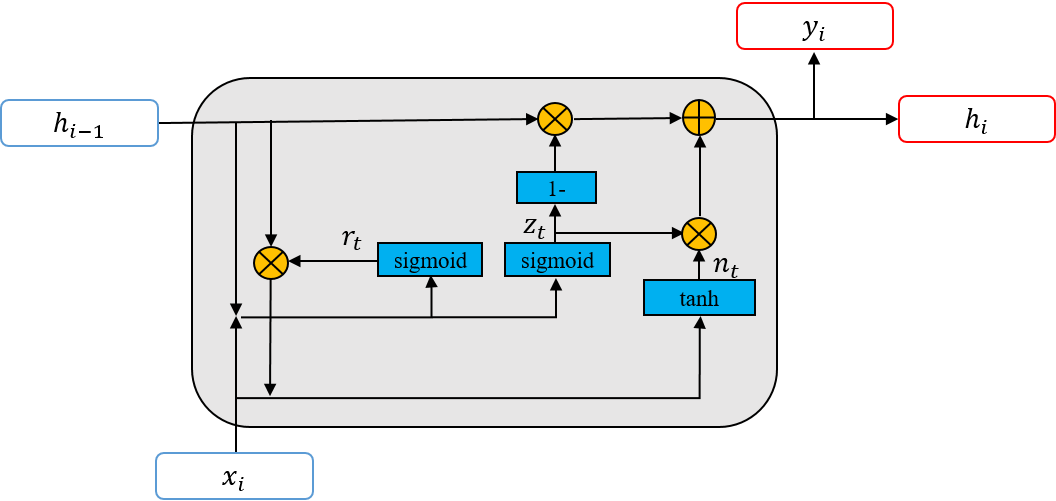}
	\caption{The illustration of gated recurrent units(GRU). $x_i$ and $y_i$ denote the current input and output of GRU,  $h_{i-1}$ and $h_i$ are the last hidden information and current hidden information, $r_i$ and $z_i$ are the reset and update gates. A GRU network is formed from a series of GRUs.}
	\label{pic:GRU}
\end{figure}

Recurrent neural network (RNN) has an excellent performance in modeling sequential data. A Gated recurrent unit (GRU) \shortcites{Cho2014}\citep{Cho2014} is an improved version of the standard recurrent neural network. It was proposed to solve the vanishing gradient problem of a standard recurrent neural network during backpropagation. The gating mechanism enables GRU to carry information from earlier time steps to later ones. The illustration of GRU is shown in Figure \ref{pic:GRU}.
In our work, GRU is used for temporal modeling.

\begin{figure}[h!]
	\centering
	\includegraphics[width = 0.4\textwidth]{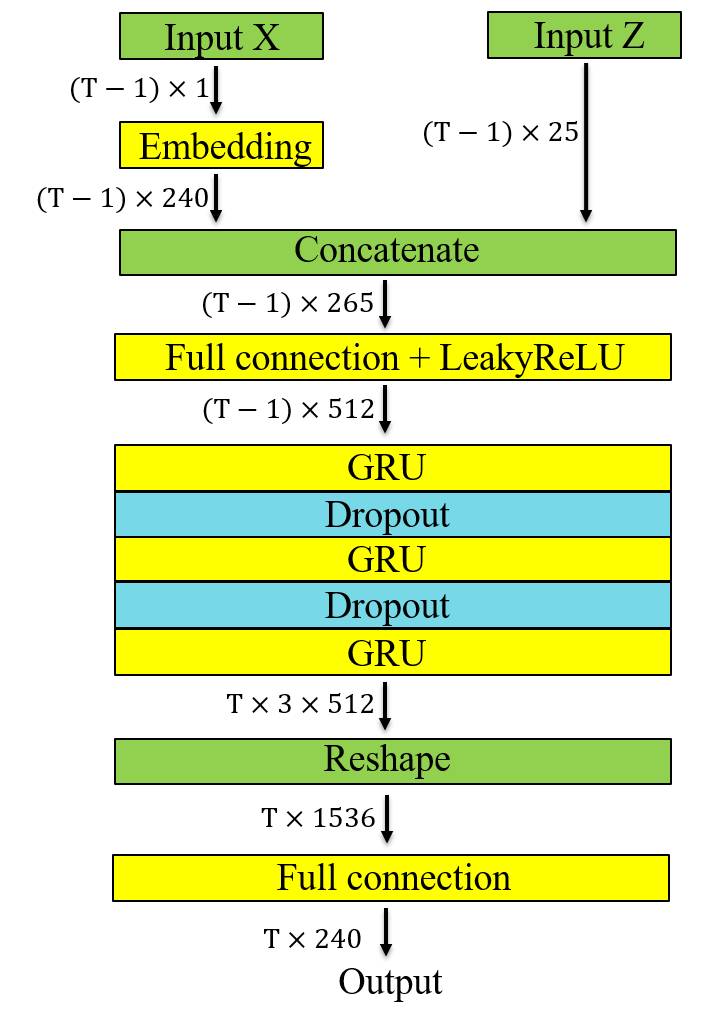}
	\caption{Diagram of the EmotionBox model architecture. "Input X" denotes a sequence of events and "Input Z" denotes the pitch histogram and note density.}
	\label{pic:Network}
\end{figure}

The model is shown in Figure \ref{pic:Network}. Input X represents the masked performance events while Input Z  represents the pitch histogram and the note density. 
Masking means the last event of each event sequence is dropped out and the rest part of the event sequence is sent to the neural network as the input. The reason for this is to make the model generate the unmasked sequence recursively. Then, we can calculate the loss, i.e. the difference, between the generated unmasked sequence and ground truth.
If the length of an event sequence is $T$, the size of Input X (i.e. the masked performance events) will be $(T-1)\times 1$. Each performance event is converted to a 240-dimension vector by a $240\times 240$ embedding layer.
The pitch histogram is a $(T-1)\times 12$ vector and note density is converted to a $(T-1)\times 12$ one-hot vector. A  $(T-1)\times 1$ zero vector is used to increase the stability of the neural network. Therefore, the size of Input Z is $(T-1)\times 25$.

The pitch histogram and note density are then concatenated with the 240-dimension vector. 
The size of the concatenated vector is $(T-1)\times 265$.
The concatenated input is fed into a $265\times 512$ full connection layer and a rectified linear unit (ReLU) activation function. Then, this $(T-1)\times 512$ vector is sent into a three-layer, 512-unit GRU, with a 0.3 dropout applied after each of the first two GRU layers. The GRU output is then fed to a 240-unit linear layer. The output of the neural network is a $T\times 240$ vector. The output presents the probability of each event at each time step.
Then, the cross-entropy loss between the generated sequence and the unmasked event sequence, namely the ground truth, is calculated.

\subsection{Emotional Music Generation}

At the generating stage, we generate samples with different emotions by specifying a particular pitch histogram and note density. When the model generating music, the first event will be randomly selected. The first event, pitch histogram, and note density are sent to the model to create new events recursively. 
The output of our model is the probability of 240 events. If we use greedy sampling to select an event with the largest probability, the sample may end up with some partial repetition, which means a small part of the music may repeat again and again.
Therefore, we combine greedy sampling with stochastic sampling. 
We select a threshold ranged from 0 to 1. Whenever a new event is sampled, we produce a random number ranged from 0 to 1. If the random number is larger than the threshold, this event will be sampled using the greedy algorithm, which means selecting an event with the largest probability. If not, this event will be sampled based on the probability of each event, which produces a lot of uncertainty. 

When generating a new piece of emotional music, we can use temperature \shortcites{He2018}\citep{He2018} to alter the degree of uncertainty. Temperature is a hyperparameter used to control the randomness of predictions by scaling the logits before applying softmax. Lower temperature results in more predictable events, while higher temperature results in more surprising events.

\section{Experiment}\label{section4}

\subsection{Dataset}

We selected a widely used dataset, piano-midi\footnote{The training data can be found on {http://www.piano-midi.de/}.}, to train our model. It includes 329 piano pieces from 23 classical composers. Each piece is a MIDI file capturing a classical piano performance with expressive dynamics and timing. The dataset is highly homogeneous because all of the pieces in it are classical music and the solo instrument is consistently piano.
The authors in \shortcites{Zhao2019}\citep{Zhao2019} labeled this dataset with four basic emotions mentioned above(i.e. happy, tensional, peaceful, and sad) manually to train their label-based automatic emotional music generator.
For the comparison experiment, we also used this emotion-labeled dataset with the permission of the authors to train a label-based model. 
The Pretty-Midi package was used to extract the note information from the MIDI files \shortcites{raffel2014intuitive}\citep{raffel2014intuitive}.

\subsection{Training}

At the training stage, the whole sequence of events is cut into 200-event-wide event sequences. The stride of event sequences is 10 events. The network was trained using the ADAM optimizer with a loss function of Cross-Entropy loss between the predicted event and ground truth event. We used a learning rate of .0002 and the model was trained for 100 epochs with a batch size of 64. We implemented our models in PyTorch.

\subsection{Comparison}

We implement a label-based model for comparison as all previous emotional music generation models were based on emotion labels \shortcites{Ferreira2019,Zhao2019}\citep{Ferreira2019,Zhao2019}. 
In order to evaluate the performance between our proposed method and the labeled-based method, the structure of the label-based model remains unchanged except that the inputs Z of the model are substituted with emotion labels. 
One-hot coding is used to present four basic emotions.
The neural network is trained to learn the mapping between music emotions and well-classified emotion labels. In the generation stage, the label-based model takes the emotion label as input.

\section{Results and Discussion}\label{section5}

To evaluate the performance of music generation given a specific emotion, subjective listening test study was carried out to compare our proposed method with the label-based method.
Similar to the human subjective listening test for analyzing different styles of classification, three 6-second long music samples were provided for each emotion and each model\footnote{The subjective listening test files can be found on {https://github.com/KaitongZheng/EmotionBoxDEMO}}.
The total amount of music samples was 24. The samples were randomly selected and shuffled. 26 subjects took part in the test.
For each sample, participants were asked which emotion they thought the sample contains. They have to choose one option from happy, peaceful, sad and tensional. It is a little difficult for untrained participants to classify the music emotion. Therefore, we provided a warming-up stage by playing four manually selected emotional music samples with their corresponding emotional labels. During the listening test, samples can be stopped, replayed to make sure the participants can hear the music clearly.

\subsection{Emotion Classification}
\begin{figure}[h!]
	\centering
	\includegraphics[width = 0.8\textwidth]{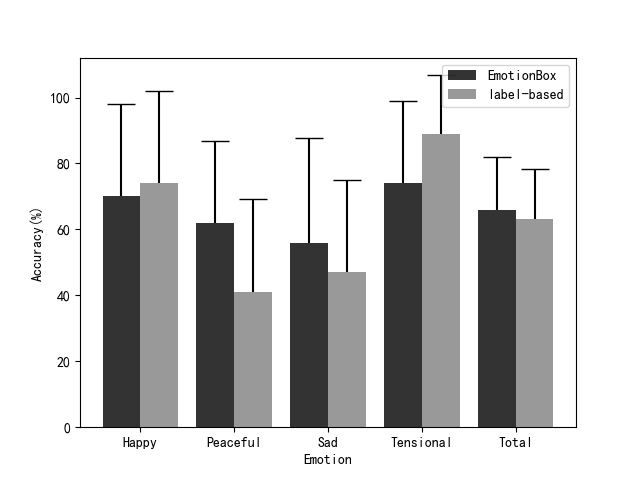}
	\caption{The mean accuracy and standard deviation of subjective evaluation test for classifying generated music samples into emotion categories.}
	\label{pic:human_test}
\end{figure}

In this section, we calculated the accuracy of emotion classification for each of four emotions and two methods. The statistical results are shown in Figure \ref{pic:human_test}. 
From Figure \ref{pic:human_test}, it shows that our proposed model, without a database labeled with emotions, has comparable performance to the label-based model in terms of emotion classification accuracy. 
Among the four kinds of emotion, the results between the two methods indicate that the music samples with a tensional emotion were correctly recognized with the highest accuracy for both methods. The music samples with a sad emotion had the lowest accuracy for our proposed method while the music samples with a peaceful emotion had the lowest accuracy for the label-based method. These observations show that emotions with higher arousal like happy and tensional are more likely to be distinguished than emotions with low arousal like sad and peaceful.
The proposed method outperforms the label-based method on peaceful and sad samples, which greatly overcomes the shortcomings of label-based method and yields a more balanced result.

Table \ref{table:wilcoxonmodel} shows a post-hoc analysis on the comparisons within each emotion pair of two methods, and the method of Wilcoxon signed-rank test is used for matched samples. 
Table \ref{table:wilcoxonmodel} indicates that there are significant differences between the two methods on tensional and peaceful samples. The emotion classification accuracy of label-based method is significantly high on tensional emotion while that is significantly low on peaceful emotion. There are no significant differences between two methods on happy and sad samples. Two methods have a similar performance on emotional music generation in general.


\begin{table}[h!]
	\centering
	\renewcommand\arraystretch{2}
	\setlength{\tabcolsep}{10pt}
	\caption{A post-hoc comparison of each emotion on their pairwise comparison between two methods, using the Wilcoxon signed-rank for matched samples $p$ value less than $0.05$ means a statistically significant difference  at a confidence level of $5\%$ and is represented in bold type.}
	\label{table:wilcoxonmodel}
	\begin{tabular}{cc c}
		\hline
		EmotionBox & label-based method & \textbf{$p$ value}\\
		\hline
		Happy & Happy & 0.4389\\
		Tensional & Tensional & \textbf{0.0026}\\
		Sad & Sad & 0.3007\\
		Peaceful & Peaceful & \textbf{0.0111}\\ 
		All & All & 0.2855\\
		\hline
	\end{tabular}
\end{table}

\begin{table}[H]
	\begin{subtable}
		\centering
		\renewcommand\arraystretch{1.3}
		\setlength{\tabcolsep}{14pt}
		\small
		\caption{The results of human classification for each combination between specific emotion at generating stage and emotion classified by subjects . (a) The results of the EmotionBox. (b) The results of the emotion-label-based model. }
		\label{table:statistical analysis}
		\caption*{(a)}
		\begin{tabular}{ |c| c| c| c| c| } 
			\hline
			\diagbox[innerwidth=4.8cm]{Generated samples}{Subjects classification} & Happy & Tensional & Sad & Peaceful \\ 
			\hline
			Happy & 71\% & 28\% & 0\% & 1\% \\
			\hline
			Tensional & 17\% & 74\% & 5\% & 4\% \\
			\hline
			Sad & 1\% & 8\% & 56\% & 35\% \\
			\hline
			Peaceful & 8\% & 4\% & 26\% & 63\% \\
			\hline
		\end{tabular}
		
		\label{tab:model1}
	\end{subtable}
	\vspace{10pt}
	\begin{subtable}
		\centering
		\renewcommand\arraystretch{1.3}
		\setlength{\tabcolsep}{14pt}
		\small
		\caption*{(b)}
		\begin{tabular}{ |c| c| c| c| c| } 
			\hline
			\diagbox[innerwidth=4.8cm]{Generated samples}{Subjects classification} & Happy & Tensional & Sad & Peaceful \\ 
			\hline
			Happy & 74\% & 23\% & 0\% & 3\% \\
			\hline
			Tensional & 10\% & 90\% & 0\% & 0\% \\
			\hline
			Sad & 4\% & 18\% & 47\% & 31\% \\
			\hline
			Peaceful & 26\% & 28\% & 5\% & 41\% \\
			\hline
		\end{tabular}
		
		\label{tab:model2}
	\end{subtable}
\end{table}

To investigate the performance of generating different emotional music within each model, we also count the result of all the combinations between specific emotion at generating stage and emotion classified by subjects as shown in Table \ref{table:statistical analysis}.
From Table \ref{table:statistical analysis}(a), it shows that the arousal of music is more distinguishable than valence. 
For example, for the first row, $28\%$ happy samples were classified as tensional samples that have the same level of arousal but a different level of valence. However, a happy sample is rarely classified as a peaceful sample as they have a different level of arousal. This experimental result agrees with the observation that tempo is more determinant than the mode in forming happy-sad judgments as reported in \shortcites{Gagnon2003}\citep{Gagnon2003}. In our work, the tempo and the mode are associated with arousal and valence of music respectively. The classification of arousal and valence will be discussed in next section.

From Table \ref{table:statistical analysis}(b), the classification accuracy is similar for high arousal music.
However, for low arousal music, the classification accuracy in terms of both arousal and valence of emotion decreases significantly. For the last row, $26\%$ and $28\%$ peaceful samples were perceived as happy samples and tensional samples respectively, which indicates that the label-based method has a poor performance on generating music with a low arousal emotion.

\subsection{Arousal and Valence Classification}

\begin{figure}[H]
	\centering
	\includegraphics[width = 0.8\textwidth]{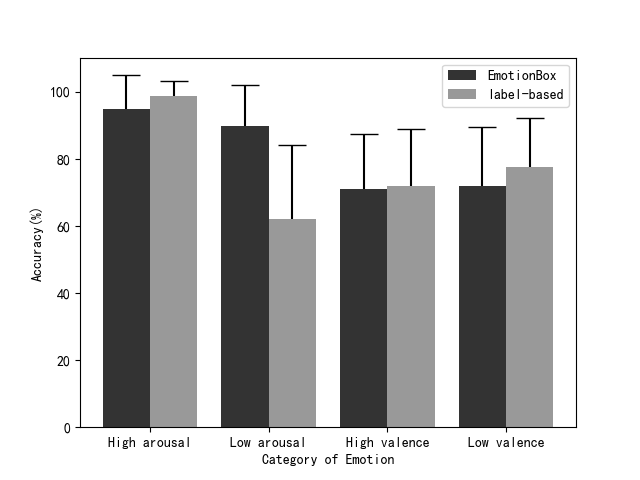}
	\caption{The mean accuracy and standard deviation of subjective evaluation test for classifying generated music samples into arousal and valence categories.}
	\label{pic:human_test_arousalvalence}
\end{figure}

Our proposed method uses note density and pitch histogram as features to present the arousal and valence of a specific emotion, respectively. To investigate whether these two features are suitable or not for training the deep neural networks, we calculated the accuracy of arousal and valence classification shown in Figure \ref{pic:human_test_arousalvalence}. 
If the emotion specified during generating stage and the emotion classified by subjects have the same arousal or valence, the classification result will be calculated as correct. For example, if the emotion of a sample specified during generating stage is happy while classified as tensional by subjects , the classification result will be viewed as correct because of the same arousal of happy and tensional.
Table \ref{table:wilcoxonarousalvalence} shows a post-hoc analysis on the comparisons between two methods in terms of arousal and valence.

\begin{table}[h!]
	\centering
	\renewcommand\arraystretch{2}
	\setlength{\tabcolsep}{10pt}
	\caption{A post-hoc comparison of each arousal and valence conditions of two methods, using the Wilcoxon signed-rank for matched samples $p$ value less than $0.05$ means a statistically significant difference  at a confidence level of $5\%$ and is represented in bold type.}
	\label{table:wilcoxonarousalvalence}
	\begin{tabular}{cc c}
		\hline
		EmotionBox & label-based method & \textbf{$p$ value}\\
		\hline
		high arousal & high arousal & 0.0794\\
		low arousal & low arousal & \textbf{8e-5}\\
		high valence & high valence & 0.961\\
		low valence & low valence & 0.1562\\
		\hline
	\end{tabular}
\end{table}

It can be observed from the results that the classification accuracy of EmotionBox is significantly higher than that of the label-based method on low arousal emotions. For other emotion categories, Table \ref{table:wilcoxonarousalvalence} shows that there is no significant difference between two methods except for low arousal emotions.
The tempo and the mode are relevant with note density and pitch histogram respectively in our work. Note density and pitch histogram further present arousal and valence respectively.
Without the limitation of note density, the label-based method tends to generate music with a faster tempo, which results in a low classification accuracy of the samples with low arousal emotions.
This result means note density is a suitable feature to control the arousal of music. 

However, the classification of valence is still challenging, which indicates that the valence of music cannot solely be presented by mode. A more appropriate presentation method of valence should be investigated in future work.

\section{Conclusion}\label{section6}
In this work, we propose a music-element-driven automatic emotional music generator. The model does not need any music datasets with emotion labels that the previous methods required. The note density and the pitch histogram are chosen to present arousal and valence of music respectively. Then, different combinations of arousal and valence will be mapped to different emotions according to the Russell emotion model.
Experimental results indicate that our proposed method has a more competitive and balanced performance in  emotional music generation than the emotion-label-based method in general. Especially, our proposed method has a significantly better performance on generating music with low arousal emotions. 
Based on the note density and pitch histogram, our proposed method will be able to generate emotional music given a specific emotion.
The results of subjective listening test indicate that note density is a suitable presentation for the arousal of music while more researches should be done to find a more appropriate feature to convey the valence of music.

\section*{Acknowledgement(s)}

We would like to thank all subjects for participating in the subjective listening test.

\section*{Funding}

This work was supported by the Open Research Project of the State Key Laboratory of Media Convergence and
Communication, Communication University of China, China (No. MCCGZ2005), and the IACAS Young Elite Researcher
Project (Grant No. QNYC201720).
This work was also partially supported by the National Science Fund of China(Grant No. 12074403 and No. 11974086), the Open Fund of National Environmental Protection Engineering and Technology Center for Road Traffic Noise Control.

\bibliographystyle{apacite}
\bibliography{MyCollection}

\end{document}